\documentclass[aps,prd,amsmath,amssymb,eqsecnum,nofootinbib]{revtex4}

\usepackage{graphicx}

\begin{document}

\title{Vacuum Polarization on the Schwarzschild Metric with a Cosmic String}
\author{Adrian C. Ottewill}
\email{adrian.ottewill@ucd.ie}
\author{Peter Taylor}
\email{peter.taylor@ucd.ie}
\affiliation{School of Mathematical Sciences and Complex \& Adaptive Systems Laboratory, University College Dublin, Belfield, Dublin 4, Ireland}

\date{\today}

\begin{abstract}
We consider the problem of the renormalization of the vacuum polarization in a symmetry space-time with axial but not spherical symmetry,  Schwarzschild space-time threaded by an infinite straight cosmic string. Unlike previous calculations, our framework to compute the renormalized vacuum polarization does not rely on special properties of Legendre functions, but rather has been developed in a way that we expect to be applicable to Kerr space-time. \end{abstract}
\maketitle

%----------------------INTRODUCTION---------------------------------------------------------------------

\section{Introduction}
\label{sec:intro}
In this paper we calculate the vacuum polarization of a massless, minimally coupled scalar field in the region exterior to the horizon of a Schwarzschild black hole threaded by an infinite thin cosmic string which reduces the spherical symmetry to axial symmetry. We consider the scalar field in the Hartle-Hawking vacuum state, corresponding to a black hole of mass $M$ in (unstable) thermal equilibrium with a bath of blackbody radiation. 

There is an extensive body of work on renormalization of the vacuum polarization on black hole spacetimes (\cite{Candelas:1980zt,Candelas:1984pg,Anderson:1989vg,Anderson:1990jh,Winstanley:2007}). In  all of these cases, the authors have considered spherically symmetric black holes. The most astrophysically significant case, however, is the Kerr-Newman black hole. The calculation of $\langle \hat{\varphi}^{2} \rangle_{ren}$ in this case has proved elusive (with the exception of it's calculation on the pole of the horizon  where the effects of rotation are minimized  \cite{Frolov:1982pi}). The principal reason for this is that the rotation of the black hole means that it no longer possesses spherical symmetry, but only axial symmetry. The existing calculations, referenced above, rely heavily on the Legendre Addition Theorem and other well known properties of the special functions, as well as the Watson-Sommerfeld formula.  In the axial symmetric case, the Legendre Addition Theorem does  not apply and the Watson-Sommerfeld formula is no longer useful
as it is no longer possible to perform the sum over the azimuthal quantum number. 

In the case being considered in this paper, we are dealing with a black hole where the symmetry is reduced from spherical to axial but without the added complication of rotation. This fact makes this case an ideal precursor to the Kerr-Newman case.
It  presents the first calculation of the renormalized vacuum polarization on the exterior region of  an axially symmetric  black hole. It should be noted that DeBenedictis~\cite{DeBenedictis} has calculated the vacuum polarization on an axially symmetric metric, but in the case of a black string, not a black hole. Most importantly, the method presented in that paper is not applicable to the current space-time or to the Kerr-Newman case.

The method we present here does not rely on specific properties of the angular functions (such as Addition Theorems) and that results in a complete mode-by-mode subtraction (as opposed to the partial mode-subtractions in the literature). Furthermore, we elucidate some important points about the Christensen-DeWitt point-splitting approach \cite{Christensen:1976vb} to renormalization. In particular, we show that the choice of point-separation direction is intimately connected to the order of summation of the mode-sum.

%-----------------------MODE-SUM EXPRESSION---------------------------------------------------------

\section{The Mode-Sum Expression for the Green's Function}
\label{sec:modesum}
The cosmic string is modeled by introducing an azimuthal deficit parameter, $\alpha$, into the standard Schwarzschild metric. We
may describe this in  coordinates $(t, r, \theta, \tilde{\phi})$, where  $\tilde{\phi}$ is periodic with period  $2\pi \alpha$, so we may take $\tilde{\phi}\in[0,  2\pi \alpha)$, in which the line element is given by
\begin{align}
\mathrm{d}s^2 = -(1-2M/r) \mathrm{d}t^2 +(1-2M/r)^{-1} \mathrm{d}r^2 \nonumber\\
+r^2 \mathrm{d} \theta^2 +  r^{2} \sin^{2}\theta \mathrm{d} \tilde{\phi}^{2}.
\end{align}
For GUT scale cosmic strings $\alpha = 1- 4 \mu$ where $\mu$ is the mass per unit length of the string so we shall assume $0 < \alpha \le 1$. In Sec.~\ref{sec:flatspacesum}, we further limit ourselves to the case $1/2 <\alpha\le 1$ which is physically justifiable since $\mu \ll 1$.
We can alternatively define a new azimuthal coordinate by
\begin{equation}
\tilde{\phi} = \alpha \phi
\end{equation}
so that  $\phi$ is periodic with period  $2\pi$ and we may take $\phi \in [0, 2 \pi)$. In coordinates $(t, r, \theta, \phi)$,  the line element is given by
\begin{align}
\mathrm{d}s^2 = -(1-2M/r) \mathrm{d}t^2 +(1-2M/r)^{-1} \mathrm{d}r^2 \nonumber\\
+r^2 \mathrm{d} \theta^2 + \alpha^{2} r^{2} \sin^{2}\theta \mathrm{d} \phi^{2}.
\end{align}

We shall consider a massless, minimally coupled scalar field, $\varphi$, in the Hartle-Hawking vacuum state. 
Since this is a thermal state, it is convenient to work with the Euclidean Green's function, performing a Wick rotation of the
 temporal coordinate $t\rightarrow -i \tau$ and eliminating the conical singularity at $r=2M$ by making $\tau$ periodic with period
$2\pi/\kappa$ where $\kappa=1/(4M)$ is  the surface gravity of the black hole. 
The massless, minimally coupled scalar field, satisfies the homogenous wave-equation
\begin{equation}
\Box \varphi(\tau,r,\theta,\phi)=0,
\end{equation}
which can be solved by a separation of variables by writing
\begin{equation}
\varphi(\tau,r,\theta,\phi)\sim e^{i n \kappa \tau+i m\phi} P(\theta)R(r)
\end{equation}
where $P(\theta)$ is regular and satisfies 
\begin{equation}
\label{eq:legendre}
\Big\{\frac{1}{\sin\theta} \frac{\mathrm{d}}{\mathrm{d} \theta}\Big(\sin\theta\frac{\mathrm{d}}{\mathrm{d} \theta}\Big) -\frac{m^{2}}{\alpha^{2}\sin^{2}\theta}+\lambda(\lambda+1)\Big\}P(\theta)=0
\end{equation}
while $R(r)$ satisfies
\begin{equation}
\label{eq:radialhomogeneous}
\Big\{\frac{\mathrm{d}}{\mathrm{d} r}(r^2-2Mr)\frac{\mathrm{d}}{\mathrm{d}r} - \lambda(\lambda+1)-\frac{n^2 \kappa^2 r^4}{r^2-2Mr}\Big\}R(r)= 0.
\end{equation}
 The $\lambda(\lambda+1)$ term arises as the separation constant. The choice of $\lambda$ is arbitrary for $\varphi$ to satisfy the wave equation but requires a specific choice in order for the mode-function $P(\theta)$ to satisfy the boundary conditions of regularity on the poles. In the Schwarzschild case (no cosmic string, $\alpha=1$), regularity on the poles means that $\lambda=l$, i.e the separation constant is $l(l+1)$. In the cosmic string case, the appropriate choice of $\lambda$ that guarantees regularity of the angular functions on the poles is
\begin{equation}
\label{eq:lambda}
\lambda=l-|m|+|m|/\alpha.
\end{equation}  
%We will, however, in the interest of being succinct, retain the notation $\lambda(\lambda+1)$ as the separation constant with the understanding that $\lambda$ is given by Eq.(\ref{eq:lambda}). 
With this choice of $\lambda$, the angular function is the Legendre function of both non-integer order and non-integer degree,\textit{viz.},
\begin{equation}
P(\theta)=P_{\lambda}^{-|m|/\alpha}(\cos\theta).
\end{equation}
It can be shown that these angular functions satisfy the following normalization condition,
\begin{align}
\label{eq:norm}
\int_{-1}^{1}&P_{l-|m|+|m|/\alpha}^{-|m|/\alpha}(\cos\theta) P_{l'-|m|+|m|/\alpha}^{-|m|/\alpha}(\cos\theta) \mathrm{d}(\cos\theta) \nonumber\\
&\qquad =\frac{2}{(2\lambda+1)}\frac{\Gamma(\lambda-|m|/\alpha+1)}{\Gamma(\lambda+|m|/\alpha+1)}\delta_{ll'}. 
\end{align}
The periodicity of the Green's function with respect to $(\tau-\tau')$ and $(\phi-\phi')$ with periodicity $2\pi/\kappa$ and $2\pi$, respectively, combined with Eq.(\ref{eq:norm}) allow us to write the mode-sum expression for the Green's function as
%\begin{widetext}
\begin{widetext}
\begin{align}
\label{eq:greensfn}
G(x,x')=\frac{T}{4\pi}\sum_{n=-\infty}^{\infty} e^{i n \kappa(\tau-\tau')}\sum_{m=-\infty}^{\infty} e^{im(\phi-\phi')} \sum_{l=|m|}^{\infty} (2\lambda+1)\frac{\Gamma(\lambda+|m|/\alpha+1)}{\Gamma(\lambda-|m|/\alpha+1)}P_{\lambda}^{-|m|/\alpha}(\cos\theta) P_{\lambda}^{-|m|/\alpha}(\cos\theta')\chi_{n\lambda}(r,r'). \nonumber\\
\end{align}
\end{widetext}
%\end{widetext}
where $\chi_{n\lambda}(r,r')$ satisfies the inhomogeneous equation,
\begin{align}
\label{eq:radialr}
\Big\{\frac{\mathrm{d}}{\mathrm{d} r}(r^2-2Mr)\frac{\mathrm{d}}{\mathrm{d}r} - &\lambda(\lambda+1)-\frac{n^2 \kappa^2 r^4}{r^2-2Mr}\Big\}\chi_{n\lambda}(r,r')\nonumber\\
&=-\frac{1}{\alpha}\delta(r-r').
\end{align}

It is convenient to write the radial equation in terms of a new radial variable
%\begin{equation}
%\label{eq:eta}
$\eta=r/M-1$,
%\end{equation}
 the radial equation then reads 
 \begin{align}
\label{eq:radial}
\Big\{\frac{\mathrm{d}}{\mathrm{d}\eta}\Big((\eta^{2}-1)\frac{\mathrm{d}}{\mathrm{d}\eta}\Big)-&\lambda(\lambda+1)-\frac{n^{2}(1+\eta)^{4}}{16(\eta^{2}-1)}\Big\}\chi_{n\lambda}(\eta, \eta')\nonumber\\
&=-\frac{1}{\alpha M}\delta(\eta-\eta').
\end{align}
where we have used the fact that $\kappa=1/(4M)$. For $n=0$, the two solutions of the homogeneous equation are the Legendre functions of the first and second kind. For $n\ne 0$, the homogeneous equation cannot be solved in terms of known functions and must be solved numerically. We denote the two solutions that are regular on the horizon and infinity (or some outer boundary) by $p_{n\lambda}(\eta)$ and $q_{n\lambda}(\eta)$, respectively. A near-horizon Frobenius analysis for $n \neq 0$ shows that the indicial exponent is $\pm |n|/2$, and so we have the following asymptotic forms:
\begin{equation}
\label{eq:asymp}
\begin{alignedat}{2}
 p_{n\lambda}(\eta)& \sim (\eta-1)^{|n|/2}\qquad\qquad &\eta\rightarrow 1 \\
 q_{n\lambda}(\eta)& \sim (\eta-1)^{-|n|/2}   &\eta \rightarrow 1.
\end{alignedat}
\end{equation}
Defining the normalizations by these asymptotic forms and using the Wronskian conditions one can obtain the appropriate normalization of the Green's function:
 \begin{equation}
 \label{eq:chi}
 \chi_{n\lambda}(\eta,\eta') = \begin{cases}
                                                  \displaystyle{ \frac{1}{\alpha M} P_{\lambda}(\eta_{<})Q_{\lambda}(\eta_{>})}&n=0  \\
                                                  \displaystyle{ \frac{1}{2|n|\alpha M}p_{n\lambda}(\eta_{<}) q_{n\lambda}(\eta_{>})}&n\ne 0.
                                                   \end{cases}
 \end{equation}

%-----------------CHOOSING A SEPARATION DIRECTION-------------------------------------------------------------------------

\section{Choosing a Separation Direction}
\label{sec:order}
In order to renormalize $\langle \hat{\varphi}^{2} \rangle$, we subtract, in a meaningful way, the geometrical Christensen-DeWitt renormalization terms from the Green's function. This approach rests on the fact that all Green's functions will possess the same  short distance behaviour 
encapsulated in the Hadamard parametrix.  On the other hand the full Green's function must reflect the relevant boundary conditions of the global problem, for example periodicity with particular period in $\tau$ and $\phi$, and this is most easily expressed by a mode decomposition.
In order to perform the renormalization it is first necessary to perform a regularization of the Green's function and
the natural regularization in the Christensen-DeWitt approach is to consider the Green's function with the two points separated.
The geometrical singularity may then be removed prior to bringing the two points together to give $\langle \hat{\varphi}^{2} \rangle_{ren}$.
The geometrical nature of the subtraction means that we obtain the same result whichever direction we separate in, although there
is a surprising twist in the tale described below.

As we are free to choose the direction of separation, we may do so in a way which makes the calculation as straightforward as possible. In almost all black hole calculations in the literature, e.g.\cite{Candelas:1984pg,JensenOttewill,Anderson:1989vg,Anderson:1990jh,Winstanley:2007}, the authors have preferred to separate in the temporal direction, except for on-horizon calculations. The principal reason for this is that the metric components do not depend on $\tau$, making the renormalization terms somewhat easier. (The same could be said for separating in $\phi$, however, making it an equally suitable candidate.) For on-horizon calculations, it is typically most convenient to separate in the radial direction (see \cite{Candelas:1980zt} for example). In fact, we have calculated elsewhere~\cite{CSHorizon} an analytic expression for $\langle \hat{\varphi}^{2} \rangle_{ren}$ on the horizon of the Schwarzschild black hole threaded by an infinite cosmic string by separating radially and using a summation formula we have derived in that paper.

Since we have calculated the vacuum polarization on the horizon in another paper, we shall concentrate here on the calculation of off-horizon values in this paper.  The question we must address is what is the most convenient separation when we no longer have spherical symmetry. 
To answer this we start by analysing the approach taken in the literature in the spherically symmetric case. In this case, one can sum over $m$ to obtain expression for $\langle \hat{\varphi}^{2} \rangle_{ren}$ that involves an inner  sum over $l$-modes and an outer sum over $n$-modes. One then converts the $l$-sum into an integral using the Watson-Sommerfeld formula and sums the $n$-modes directly. (This is, of course, only a sketch of the method and several other tricks and techniques are used to do the calculation, none of which are important to the choice of separation.) 

In the cosmic string case, the axial symmetry means that converting an $l$-sum to an integral results in numerical integrals over the square of Legendre functions; this is neither convenient nor useful.
The most practical way to proceed with the calculation is to convert the $n$-sum to an integral and sum the $l$, $m$-modes directly, since the angular part of the Green's function does not depend on the $n$-modes. In fact, this proves to be a very fruitful approach to these calculations, both in the axially symmetric and the spherically symmetric case. There is a caveat, however,
which relates to the order in which we perform the mode sums which is intimately related to the distributional nature of the expressions we are dealing with.

%  One might naively assume that the order of these sums shouldn't matter since the Killing vectors $\partial/\partial\tau$ and $\partial/\partial\phi$ commute. This means that, when we derive our Green's function, it matters not whether we separate out the $\tau$ dependence or the $\phi$ dependence first. But, in some sense, this symmetry is being broken by choosing a preferred direction in which to separate. 

At a 4-dimensional level we have a sum over a complete set of mode functions and the expression
\[
G(x,x')= \sum_i \frac{u_i(x)u_i(x')}{\lambda_i}
\]
is understood in the sense of smearing with smooth functions of compact support in the 4-dimensional space.
In moving to a point separated expression with just one coordinate different we must consider the $\delta$ convergent
limits in three of our coordinate directions.  In particular as  $\partial/\partial\tau$ and $\partial/\partial\phi$ are commuting Killing vectors we may associate with them independent quantum numbers $n$ and $m$. Also correspondingly, when we derive our Green's function, it matters not whether we separate out the $\tau$ dependence or the $\phi$ dependence first. However,
when we limit our test functions to 3-dimensional delta functions the corresponding sum must remain until last,
i.e., the outer sum must be that which corresponds to the direction in which we have separated. More specifically, separating in the temporal direction corresponds to an inner $l$, $m$-sum and an outer $n$-sum, separating in the azimuthal direction corresponds to an inner $n$-sum and an outer $l$, $m$-sum.

We have two strong argumets in support of the above argumet.  Firstly, one can show analytically, in the Schwarzschild case, that the finite difference between summing in the different orders is precisely the analytic difference between the regular parts of the Christensen subtraction terms for temporal and azimuthal separation. In other words, separating in the temporal direction and doing the $n$-sum first gives a finite but wrong answer!  
An alternative way of understand this is that  the renormalization procedure adopted by Candelas-Howard and the alternative procedure of this paper results in a double mode-sum that is convergent, but not absolutely convergent. As a result the order of summation matters.

%One might naturally ask that if we get two different finite answers depending on the order of summation, how do we know that doing the sum associated with the separation direction last is right. Apart from the fact that this seems like the natural way to do the sums, 
The second supporting argument is that we have an unambiguous answer on the horizon in both the Schwarzschild black hole with a cosmic string \cite{CSHorizon} and without a string \cite{Candelas:1980zt}. Clearly the  correct off-horizon answer must match up with the horizon value as the horizon is approached and  this is only true when we sum according to the rules laid out above. 

It is natural for us to separate in the azimuthal direction for the cosmic string case and so in light of the previous arguments,  the appropriate mode-sum form for our unrenormalized Green's function  is
\begin{widetext}
\begin{equation}
\label{eq:gunren}
G(r,\theta,\Delta\phi)=\frac{T}{4\pi \alpha} \sum_{m=-\infty}^{\infty}e^{im\Delta\phi} \sum_{l=|m|}^{\infty}(2\lambda+1)\frac{\Gamma(\lambda+|m|/\alpha+1)}{\Gamma(\lambda-|m|/\alpha+1)} P_{\lambda}^{-|m|/\alpha}(\cos\theta)^{2} \sum_{n=-\infty}^{\infty}\chi_{n\lambda}(\eta,\eta),
\end{equation}
taking $\Delta\phi=\phi-\phi'$.
\end{widetext}

%---------------------RENORMALIZATION---------------------------------------------------------------------------------

\section{Renormalization}
For a massless scalar field in a Ricci-flat space-time, the only Christensen-DeWitt subtraction term
required for the calculation of $\langle \hat{\varphi}^2 \rangle_{ren}$ is
\begin{equation}
G_{div}(x,x')=\frac{1}{8\pi^{2} \sigma(x,x')},
\end{equation}
where $2 \sigma(x,x')$ is the square of the geodesic distance between $x$ and $x'$ \cite{Christensen:1976vb}. For an azimuthal splitting this becomes
%\begin{widetext}
\begin{align}
\label{eq:gdiv}
&G_{div}(r,\theta,\Delta\phi)=\frac{1}{4\pi^{2}}\Bigl[\frac{1}{g_{\phi\phi}\Delta\phi^{2}}+\frac{g_{a b}\Gamma_{\phi\phi}^{a}\Gamma_{\phi\phi}^{b}}{12 g_{\phi\phi}^{2}} +O(\Delta\phi)\Bigr] \nonumber\\
&\quad=\frac{1}{4\pi^{2}M^{2}}\Bigl[\frac{1}{(\eta+1)^{2}\sin^{2}\theta\alpha^{2}\Delta\phi^{2}}+\nonumber\\
&\qquad\quad\frac{1}{12}\frac{1}{(\eta+1)^{2}\sin^{2}\theta}-\frac{1}{6}\frac{1}{(\eta+1)^{3}}+O(\Delta\phi)\Bigr] .
\end{align}
%\end{widetext}
One is now faced with the challenge of subtracting this geometrical expression from the mode sum Eq.(\ref{eq:gunren}) in such a way that the limit may be performed. The approach in the literature, e.g.~\cite{Candelas:1984pg,Anderson:1989vg,Anderson:1990jh,Winstanley:2007}, is to bring the divergent term inside the outer sum (in this case the $m$-sum, in the temporal splitting case the $n$-sum) using an identity  from distribution theory,
\begin{equation}
\frac{1}{\Delta\phi^{2}}=-\sum_{m=1}^{\infty} m e^{i m\Delta\phi}-\frac{1}{12}+O(\Delta\phi)^{2}
\end{equation}
or its equivalent expression for temporal separation. The success of this approach in the spherically symmetric case relies heavily on the applicability of the Legendre Addition Theorem and the Watson-Sommerfeld formula. This is not possible in the general case. Instead we would like to be able to write the divergent term in $G_{div}$ as a triple mode-sum over $m$, $l$,
$n$ so that a full mode-by-mode subtraction may be performed.

It turns out that there is a natural and general way to approach this problem. There are a whole gamut of summation formulae that can be derived simply by equating different but equivalent expressions for the same Green's function. This requires little knowledge of the special functions being summed, only that they are solutions to the homogeneous wave equation being considered. Indeed, this approach is  one of the standard ways of proving the Legendre Addition Theorem. 
The universality of the geometrical singularity structure provided by the Hadamard form then ensures, that with appropriate parameterisation, we can match the required coordinate divergence to an appropriate mode sum. 

Ideally, we equate a mode-sum expression for a Green's function to a closed-form or quasi-closed form expression.
The most effective way of deriving such summation formulae is by considering appropriate Green's functions on Minkowski spacetime, where  the Green's function is usually known in closed form or quasi-closed form.
%There are two good intuitive reasons for this: Firstly, when we zoom in on any spacetime geometry, it 'looks like' flat space. %This means that the same divergence that inflicts flat space will also inflict Schwarzschild (or any other) spacetime.
%Secondly, in Minkowski spacetime, the Green's function is usually known in closed form or quasi-closed form.
In the next section, we will derive the appropriate summation formula for the current problem  by considering the thermal Green's function on Minkowski spacetime threaded by an infinite cosmic string. 
%This is clearly the natural case to consider given it's analogy to the Green's function on the Schwarzschild spacetime with a cosmic string. 

%---------------THERMAL GREENS FUNCTION ON FLAT SPACE------------------------------------------------------------------

\section{Thermal Green's Function on Minkowski Spacetime with a Cosmic String}
\label{sec:flatspacesum}
We consider a massless scalar field at non-zero temperature $T$ propagating in Minkowski spacetime with an infinite cosmic string running along the polar axis. The Euclideanized metric is given by
\begin{equation}
\mathrm{d}s^{2} = \mathrm{d}\tau^{2} + \mathrm{d}\xi^{2} + \xi^{2}\mathrm{d}\theta^{2} + \alpha^{2}\xi^{2}\sin^{2}\theta \mathrm{d}\phi^{2},
\end{equation}
where $\theta$ and $\phi$ are the usual polar coordinates on the 2-sphere and $\xi$ is an arbitrary radial variable.
The periodic (thermal) Euclidean Green's function $G_\beta(\Delta\tau , \Delta\mathbf{x})$ can be written as an image sum over the zero temperature Green's function, %(see \cite{BD} for example). If 
$G(\Delta\tau , \Delta\mathbf{x}')$
% is the zer,o temperatue Green's function, then the thermal Green's function is given by
as
\begin{equation}
G_\beta(\Delta\tau , \Delta\mathbf{x})=\sum_{k=-\infty}^{\infty} G(\Delta\tau+k\beta, \Delta\mathbf{x})
\end{equation}
where $\beta=1/T$ is the inverse temperature. A mode-sum expression for the scalar field Green's function at zero temperature is easily found to be
\begin{widetext}
\begin{align}
G(\Delta\tau, \Delta\mathbf{x}) &=\frac{1}{8\pi^{2}\alpha}\int_{0}^{\infty}e^{i\omega \Delta\tau} \mathrm{d}\omega \sum_{m=-\infty}^{\infty}e^{im\Delta\phi} \sum_{l=|m|}^{\infty}(2 \lambda +1)\frac{\Gamma( \lambda+|m|/\alpha+1)}{\Gamma(\lambda-|m|/\alpha+1)}  \nonumber\\
&\qquad\qquad P^{-|m|/\alpha}_{\lambda}(\cos\theta)P^{-|m|/\alpha}_{\lambda}(\cos\theta') \frac{1}{(\xi\xi')^{1/2}} I_{\lambda+1/2}(\omega \xi_{<}) K_{\lambda+1/2}(\omega \xi_{>})
\end{align}
\end{widetext}
where $I$ and $K$ are the modified Bessel functions \cite{gradriz} of the first and second kind, respectively. 
%Therefore the thermal Green's function is given by
%\begin{align}
%G_\beta(\Delta\tau;\Delta\mathbf{x}) &=\frac{1}{8\pi^{2}\alpha}\int_{0}^{\infty}\sum_{k=-\infty}^{\infty}e^{i\omega (\tau-\tau'+k\beta)} d\omega \sum_{m=-\infty}^{\infty} e^{im\Delta\phi} \sum_{l=|m|}^{\infty}(2 \lambda +1)\frac{\Gamma( \lambda+|m|/\alpha+1)}{\Gamma(\lambda-|m|/\alpha+1)}  \nonumber\\
%&&\,\,\,\,\, \,\,\,\,\,\,\,\,\,\,\,\,\,\,\,\,\times P^{-|m|/\alpha}_{\lambda}(\cos\theta)P^{-|m|/\alpha}_{\lambda}(\cos\theta') \frac{1}{(\xi\xi')^{1/2}} I_{\lambda+1/2}(\omega \xi_{<}) K_{\lambda+1/2}(\omega \xi_{>})
%\end{align}
Now, using the Fourier transform of a Comb function, 
%\begin{align}
%\sum_{k=-\infty}^{\infty} e^{i \omega(\tau-\tau'+k\beta)} &= e^{i\omega(\tau-\tau')}\sum_{k=-\infty}^{\infty}e^{i\omega k\beta} \nonumber\\
%&= e^{i\omega(\tau-\tau')}\frac{2\pi}{\beta}\sum_{n=-\infty}^{\infty}\delta(\omega-n 2\pi/\beta),
%\end{align}
\begin{align}
\sum_{k=-\infty}^{\infty} e^{i \omega k\beta} = \frac{2\pi}{\beta}\sum_{n=-\infty}^{\infty}\delta(\omega-n 2\pi/\beta),
\end{align}
we arrive at an appropriate expression for the thermal Green's function:
\begin{widetext}
\begin{align}
G_\beta(x,x') &=\frac{T}{4\pi\alpha}\sum_{n=-\infty}^{\infty}e^{i n \kappa \Delta\tau} \sum_{m=-\infty}^{\infty} e^{im\Delta\phi} \sum_{l=|m|}^{\infty}(2 \lambda +1)\frac{\Gamma( \lambda+|m|/\alpha+1)}{\Gamma(\lambda-|m|/\alpha+1)}  \nonumber\\
&\qquad\qquad P^{-|m|/\alpha}_{\lambda}(\cos\theta)P^{-|m|/\alpha}_{\lambda}(\cos\theta') \frac{1}{(\xi\xi')^{1/2}} I_{\lambda+1/2}(n\kappa \xi_{<}) K_{\lambda+1/2}(n\kappa \xi_{>})
\end{align}
where $\kappa = 2 \pi T$, and the $n=0$ term is understood in the sense 
\begin{equation}
\lim_{n\rightarrow 0} (\xi\xi')^{-1/2} I_{\lambda+1/2}(n\kappa \xi _{<}) K_{\lambda+1/2}(n\kappa \xi _{>})=\Big(\frac{\xi_{<}}{\xi_{>}}\Big)^\lambda\frac{1}{\xi_{>}(2\lambda+1)}.
\end{equation}
In particular, for azimuthal separation, we have (taking into account that we must choose the appropriate order of summation)
\begin{align}
\label{eq:flatmodesum}
G_\beta(r,\theta,\Delta\phi) &=\frac{T}{4\pi\alpha}\sum_{m=-\infty}^{\infty}e^{i m\Delta\phi} \sum_{l=|m|}^{\infty}(2 \lambda +1)\frac{\Gamma( \lambda+|m|/\alpha+1)}{\Gamma(\lambda-|m|/\alpha+1)} \nonumber\\ &\qquad\qquad P^{-|m|/\alpha}_{\lambda}(\cos\theta)^{2} \sum_{n=-\infty}^{\infty}  \frac{1}{\xi} I_{\lambda+1/2}(n\kappa \xi) K_{\lambda+1/2}(n\kappa \xi)
\end{align}
Note that, with the exception of the radial functions, this is completely equivalent to the Green's function Eq~(\ref{eq:gunren}).

We have an equivalent expression for this Green's function given by Linet~\cite{LinetCosmicString}, valid for $1/2<\alpha\le1$, in which the author writes the Green's function as the singular part plus a regular integral part, 
%\begin{equation}
$G_\beta(x,x') = G_{sing}(x,x') + G_{int}(x,x')$,
%\end{equation}
The integral part $G_{int}$ is given by
\begin{equation}
\label{eq:linetintdef}
G_{int}(x,x') = -f(\Delta\phi+\pi/\alpha)+f(\Delta\phi-\pi/\alpha)
\end{equation}
where
\begin{equation}
\label{eq:linetint}
f(\Psi)=\frac{T \sin(\Psi)}{8 \pi^{2}\alpha} \int_{0}^{\infty} \frac{\sinh(\kappa R(u))}{R(u)[\cosh(\kappa R(u)) - \cos \kappa\Delta\tau]}\frac{1}{(\cosh(u/\alpha)-\cos(\Psi))} \mathrm{d}u
\end{equation}
and
\begin{equation}
R(u) = [\xi^{2}+\xi'^{2}-2\xi\xi'\cos\theta\cos\theta'+2\xi\xi'\sin\theta\sin\theta'\cosh u]^{1/2}.
\end{equation}
The singular part is
\begin{equation}
G_{sing}(x,x') = \frac{T}{4\pi} \frac{\sinh\kappa\rho}{\rho[\cosh\kappa\rho-\cos \kappa\Delta\tau]}
\end{equation}
where
%\begin{equation}
$\rho=[\xi^{2}+\xi'^{2}-2\xi\xi'\cos\theta\cos\theta'-2\xi\xi'\sin\theta\sin\theta'\cos(\alpha\Delta\phi)]^{1/2}$.
%\end{equation}

In particular, for azimuthal point separation, we have
\begin{equation}
\label{eq:flatlinet}
G_\beta(\xi,\theta,\Delta\phi)=\frac{1}{4\pi^{2}\xi^{2}\alpha^{2}\sin^{2}\theta \Delta\phi^{2}}+\frac{1}{48\pi^{2}\xi^{2}\sin^{2}\theta}+\frac{T^{2}}{12}+G_{int}(\xi,\theta,\Delta\phi)+O(\Delta\phi)^{2}.
\end{equation}
 
 Finally, equating Eqs.~(\ref{eq:flatmodesum}) and (\ref{eq:flatlinet}), we arrive at the very useful identity
 \begin{align}
 \label{eq:triplesumformula}
 \frac{1}{4\pi^{2}}\frac{1}{\alpha^{2}\sin^{2}\theta\Delta\phi^{2}} &= \frac{T}{4\pi\alpha}\Big\{\sum_{m=-\infty}^{\infty}e^{i m\Delta\phi} \sum_{l=|m|}^{\infty}(2 \lambda +1)\frac{\Gamma( \lambda+|m|/\alpha+1)}{\Gamma(\lambda-|m|/\alpha+1)}  P^{-|m|/\alpha}_{\lambda}(\cos\theta)^{2} \nonumber\\
 & \sum_{n=-\infty}^{\infty}  \xi I_{\lambda+1/2}(n\kappa \xi) K_{\lambda+1/2}(n\kappa \xi) \Big\}- \frac{1}{48\pi^{2}\sin^{2}\theta}-\frac{T^{2}\xi^{2}}{12}-\xi^{2}G_{int}(\xi,\theta,\Delta\phi) + O(\Delta\phi)^{2}.
\end{align}
\end{widetext}
%Although this equality has been derived in the framework of Green's functions, it's veracity is independent of the framework in which it was derived. In fact we are completely free to choose the temperature $T$ and the radial variable $\xi$. The obvious choice for $T$ is the temperature of the Schwarzschild Black Hole so that $\kappa$ is now the Schwarzschild surface gravity $\kappa = 1/4M$. The choice for $\xi$ is less obvious, but we shall see in the next section that there is in fact a prescription to assign $\xi$ in terms of the Schwarzschild radial variable $\eta$.
It is important to emphasize that this equation is true for any $T$ and for all $\xi$. For our purposes, the obvious choice for $T$ is the temperature of the Schwarzschild black hole so that $\kappa$ is now the Schwarzschild surface gravity $\kappa = 1/(4M)$. In the next section we show that there is a prescription to assign $\xi$ in terms of the Schwarzschild radial variable $\eta$ in such a way as to guarantee the convergence of the mode-sum.

%----------------MODE BY MODE SUBTRACTION---------------------------------------------

\section{Mode By Mode Subtraction}
\label{sec:modesubtraction}
We now return to the Schwarzschild cosmic string case. From Eq.~(\ref{eq:gunren}) and Eq.~(\ref{eq:gdiv}), we have the following expression for the renormalized vacuum polarization:
\begin{widetext}
\begin{align}
\label{eq:gminusgdiv}
\langle \hat{\varphi}^{2} \rangle_{ren}&= \lim_{\Delta\phi\rightarrow 0} [G(r,\theta,\Delta\phi)-G_{div}(r,\theta,\Delta\phi)] \nonumber\\
&=\lim_{\Delta\phi\rightarrow 0} \Big[\frac{T}{4\pi \alpha} \sum_{m=-\infty}^{\infty}e^{im\Delta\phi} \sum_{l=|m|}^{\infty}(2\lambda+1)\frac{\Gamma(\lambda+|m|/\alpha+1)}{\Gamma(\lambda-|m|/\alpha+1)} P_{\lambda}^{-|m|/\alpha}(\cos\theta)^{2} \sum_{n=-\infty}^{\infty}\chi_{n\lambda}(\eta,\eta) \nonumber\\
&-\frac{1}{4\pi^{2}M^{2}}\frac{1}{(\eta+1)^{2}\alpha^{2}\Delta\phi^{2}\sin^{2}\theta}-\frac{1}{48\pi^{2}M^{2}}\frac{1}{(\eta+1)^{2}\sin^{2}\theta}+\frac{1}{48\pi^{2}M^{2}}\frac{2}{(\eta+1)^{3}}+O(\Delta\phi^{2})\Big]
\end{align}
\end{widetext}
%It is clear that taking the limit is a meaningless operation in the form that the expression is written. However, comparing the formula derived in the previous section Eq.(\ref{eq:triplesumformula}), we see that we have precisely the correct divergence as that above.
Now our identity, Eq.(\ref{eq:triplesumformula}), is of precisely the correct form to allow us to convert the $1/\Delta\phi^{2}$ term into an appropriate triple mode-sum. On dividing Eq.~(\ref{eq:triplesumformula}) by $M^{2} (\eta+1)^{2}$ and substituting into Eq.~(\ref{eq:gminusgdiv}), we obtain
\begin{widetext}
\begin{align}
&\langle \hat{\varphi}^{2} \rangle_{ren} = \lim_{\Delta\phi\rightarrow 0}\Big[\frac{T}{4\pi \alpha} \sum_{m=-\infty}^{\infty}e^{im\Delta\phi} \sum_{l=|m|}^{\infty}(2\lambda+1)\frac{\Gamma(\lambda+|m|/\alpha+1)}{\Gamma(\lambda-|m|/\alpha+1)} P_{\lambda}^{-|m|/\alpha}(\cos\theta)^{2} \nonumber\\
&\qquad\qquad\qquad\qquad\sum_{n=-\infty}^{\infty}\Big\{\chi_{n\lambda}(r,r) 
-\frac{\xi}{M^{2} (\eta+1)^{2}} I_{\lambda+1/2}(n\kappa \xi) K_{\lambda+1/2}(n\kappa \xi) \Big\}\nonumber\\
&\qquad\qquad+\frac{T^{2}}{12}\frac{\xi^{2}}{M^{2}(\eta+1)^{2}}+\frac{1}{48\pi^{2}M^{2}}\frac{2}{(\eta+1)^{3}} +\frac{\xi^{2}}{M^{2}(\eta+1)^{2}}G_{int}(\xi,\theta,\Delta\phi) + O(\Delta\phi^{2})\Big].
\end{align}

We can now take the limit inside the sum to get 
\begin{equation}
\langle \hat{\varphi}^{2} \rangle_{ren}=\langle \hat{\varphi}^{2} \rangle_{sum}+\langle \hat{\varphi}^{2} \rangle_{int}+\langle \hat{\varphi}^{2} \rangle_{analytic},
\end{equation}
where
\begin{align}
\label{eq:phi2sum}
&\langle \hat{\varphi}^{2} \rangle_{sum}=\frac{T}{2\pi \alpha}  \sum_{m=-\infty}^{\infty}\sum_{l=|m|}^{\infty}(2\lambda+1)\frac{\Gamma(\lambda+|m|/\alpha+1)}{\Gamma(\lambda-|m|/\alpha+1)} P_{\lambda}^{-|m|/\alpha}(\cos\theta)^{2} \nonumber\\
&\qquad \Big[ \sum_{n=1}^{\infty}\Big\{\frac{p_{n\lambda}(\eta) q_{n\lambda}(\eta)}{2 |n| M} -\frac{\xi I_{\lambda+1/2}(n\kappa \xi) K_{\lambda+1/2}(n\kappa \xi)}{M^{2} (\eta+1)^{2}}  \Big\} +\frac{1}{2}\Big\{\frac{1}{M}P_{\lambda}(\eta) Q_{\lambda}(\eta)-\frac{\xi}{M^{2} (\eta+1)^{2}(2\lambda+1)} \Big\}\Big]\\
%\end{align}
%\begin{align}
 \label{eq:phi2int}
&\langle \hat{\varphi}^{2} \rangle_{int}= -\frac{T \sin(\pi/\alpha)}{4 \pi^{2} M^{2} \alpha} \frac{\xi^{2}}{(\eta+1)^{2}}\int_{0}^{\infty}\frac{\sinh(\kappa R(u))}{R(u)(\cosh(\kappa R(u))-1)(\cosh(u/\alpha)-\cos(\pi/\alpha))} \mathrm{d}u \\
\label{eq:phi2analytic}
&\langle \hat{\varphi}^{2} \rangle_{analytic}=\frac{T^{2}}{12}\frac{\xi^{2}}{M^{2}(\eta+1)^{2}}+\frac{1}{48\pi^{2}M^{2}}\frac{2}{(\eta+1)^{3}}
\end{align}
\end{widetext}
where $R(u)=(2 \xi^{2} \sin^{2}\theta (1+\cosh u))^{1/2}$. We have used Eqs.~(\ref{eq:linetintdef}) and (\ref{eq:linetint}) to obtain the expression for $\langle \hat{\varphi}^{2} \rangle_{int}$. We have also explicitly separated out the $n=0$ term.

%This method elucidates the fact that renormalization corresponds to subtracting the flat space contribution since we have subtracted the divergent part of the flat space thermal Green's function, scaled by the geometry of the Schwarzschild spacetime. 
%This feels somewhat analagous to the normal ordering operation of QFT on flat space, where we subtract the divergent contribution from the 'vacuum'. 
A key feature of our approach is that since we have a triple mode-by-mode subtraction, with the correct choice of $\xi$, we can ensure the convergence of the triple sum, thus making redundant the removing of the 'superficial' divergence discussed by some authors (\cite{Winstanley:2007,Anderson:1989vg}). Specifically, we choose $\xi$, such that, for a given $n$, the $l$, $m$ mode-sum is regular. We can do this by associating the $l$, $m$ mode-sum with a particular 3D Green's function by a technique known as dimensional reduction \cite{Winstanley:2007}, and then we examine the Hadamard singularity structure of this Green's function. In the Appendix, we show that the condition of regularity of the $l$, $m$ sum
can be achieved by taking $\xi$ to be
\begin{equation}
 \xi = M\frac{(\eta+1)^{2}}{(\eta^{2}-1)^{1/2}}.
 \end{equation}
The mode-sum now converges for this choice of $\xi$, as we discuss in detail in the following section.
 
 %-------------------------THE WKB APPROXIMATION-------------------------------------------------------------------
 
\section{WKB Approximations}
In order to analyze the convergence of the mode-sum of Eq.~(\ref{eq:phi2sum}), we consider the WKB approximations of both the radial part of the Green's function and the subtraction terms. 
 We have adapted the WKB prescription given by Howard \cite{Candelas:1984pg, Winstanley:2007}, valid for large $n$ and $l$. Employing a second order WKB approximation, we have
 \begin{equation}
 \frac{1}{2|n|M} p_{n\lambda}(\eta)q_{n\lambda}(\eta) \sim \beta^{(0)}_{n\lambda}+\beta^{(1)}_{n\lambda}+\beta^{(2)}_{n\lambda}
 \end{equation}
 where $ \beta^{(0)}_{n\lambda}$, $ \beta^{(1)}_{n\lambda}$ and $ \beta^{(2)}_{n\lambda}$ are the zeroth, first and second order approximants respectively. Each term is written in successive powers of $\Psi_{n\lambda}(\eta)^{-1}$,
 where
 \begin{equation}
 \label{eq:psiwkb}
 \Psi_{n\lambda} = \Big(((\lambda+1/2)^{2}(\eta^{2}-1)+\omega_{n}^{2}\Big)^{1/2}
 \end{equation}
and $\omega_{n}=(n/4) (\eta+1)^{2}$. In terms of $\Psi_{n\lambda}$, the approximants are
 \begin{widetext}
 \begin{align}
 \label{eq;wkb}
 \beta^{(0)}_{n\lambda} &=\frac{1}{2 M \Psi_{n\lambda}},  \qquad
 \beta^{(1)}_{n\lambda} =\frac{1}{16 M \Psi_{n\lambda}^{3}}-\frac{\omega_{n}^{2}}{8 M \Psi_{n\lambda}^{5}}(2\eta^{2}-6\eta+7)+\frac{5\omega_{n}^{4}}{16 M\Psi_{n\lambda}^{7}}(\eta-2)^{2}, \nonumber\\
 \beta^{(2)}_{n \lambda} &= \frac{11+16\eta^{2}}{256 M \Psi_{n\lambda}^{5}}+\frac{\omega_{n}^{2}}{64 M \Psi_{n\lambda}^{7}}(-171+70\eta-88\eta^{2}+60\eta^{3}-16\eta^{2}) + \frac{7\omega_{n}^{4}}{128 M\Psi_{n\lambda}^{9}}(666-1020\eta +773\eta^{2}-320\eta^{3}+56\eta^{4})  \nonumber\\
 &- \frac{231\omega_{n}^{6}}{64 M\Psi_{n\lambda}^{11}}(\eta-2)^{2}(7-6\eta+2\eta^{2}) + \frac{1155\omega_{n}^{8}}{256 M\Psi_{n\lambda}^{13}} (\eta-2)^{4} .
 \end{align}
 We also require the WKB approximation to the subtraction term,
 \begin{align}
 \frac{\xi}{M^{2}(\eta+1)^{2}} I_{\lambda+1/2}(n \kappa \xi)K_{\lambda+1/2}(n \kappa \xi) \sim \gamma^{(0)}_{n\lambda}+\gamma^{(1)}_{n\lambda}+\gamma^{(2)}_{n\lambda}
 \end{align}
 where
 \begin{align}
 \gamma^{(0)}_{n\lambda}&=\frac{1}{2 M\Psi_{n\lambda}}  ,  \qquad
 \gamma^{(1)}_{n\lambda}=-\frac{1}{ 4 M}\frac{\omega_{n}^{2}}{\Psi_{n\lambda}^{5}} (\eta^{2}-1) +\frac{5}{16 M}\frac{\omega_{n}^{4}}{\Psi_{n\lambda}^{7}}(\eta^{2}-1), \nonumber\\
 \gamma^{(2)}_{n\lambda}&=-\frac{1}{4 M}\frac{\omega_{n}^{2}}{\Psi_{n\lambda}^{7}}(\eta^{2}-1)^{2}+\frac{49}{16 M}\frac{\omega_{n}^{4}}{\Psi_{n\lambda}^{9}}(\eta^{2}-1)^{2} -\frac{231}{32 M}\frac{\omega_{n}^{6}}{\Psi_{n\lambda}^{11}}(\eta^{2}-1)^{2}+\frac{1155}{256 M}\frac{\omega_{n}^{8}}{\Psi_{n\lambda}^{13}} (\eta^{2}-1)^{2}.
  \end{align}
\end{widetext}

Immediately, we see that the zeroth order terms are equal and therefore the slowest order term in the mode-sum is proportional to $\Psi_{n\lambda}^{-3}$. From Eq.(\ref{eq:psiwkb}), this implies that for large $l$ and $n$, the summand is $O(l/(l^{2}+n^{2})^{3/2})$, since the angular functions scale linearly with $l$. Thus, it is now clear from the cancellation of the zeroth order approximations that the mode-sum converges. This proof of convergence is considerably simpler than  analagous proofs in the standard approach (e.g., \cite{Candelas:1984pg, Winstanley:2007} ). However, though we have shown that the mode-sum converges, the convergence is extremely slow (as is usually the case with such calculations) and is not absolute as discussed in Sec.~\ref{sec:order}. 
%We have already dealt with the fact that the convergence is not absolute and this is not an issue so long as we sum in the correct order.

A standard trick for speeding the convergence in order to make the mode-sum calculation amenable is to subtract and add the second order WKB approximations \cite{Candelas:1984pg}. For the $n=0$ terms, the subtraction term is exactly the zeroth order approximant to the radial Green's function and so we need only subtract and add $\beta^{(1)}_{0\lambda}$ and $\beta^{(2)}_{0\lambda}$ terms in this case. We now have the following expression for the mode-sum
\begin{equation}
\label{eq:phi2sumwkb}
\langle \hat{\varphi}^{2} \rangle_{sum}=\frac{T}{2\pi \alpha}\Big(\Sigma +\Lambda\Big)
\end{equation}
where $\Lambda$ is the slowly converging part of the mode-sum. We further write
\begin{equation}
\label{eq:sigma}
\Sigma=\Sigma_{1}-\Sigma_{2}+1/2\Sigma_{3}
\end{equation}
where
\begin{widetext}
\begin{align}
\label{eq:sigmaterms}
  \Sigma_{1}&=\sum_{n=1}^{\infty}\sum_{l=0}^{\infty}
 \sum_{m=-l}^{l}(2 \lambda +1)\frac{\Gamma( \lambda+|m|/\alpha+1)}{\Gamma(\lambda-|m|/\alpha+1)}P^{-|m|/\alpha}_{\lambda}(\cos\theta)^{2} \Big\{\frac{ p_{n\lambda}(\eta)q_{n\lambda}(\eta)}{2|n|M}-\beta^{(0)}_{n\lambda}-\beta^{(1)}_{n\lambda}-\beta^{(2)}_{n\lambda}\Big\}  \nonumber\\
 \Sigma_{2}&=\sum_{n=1}^{\infty}\sum_{l=0}^{\infty}
 \sum_{m=-l}^{l}(2 \lambda +1)\frac{\Gamma( \lambda+|m|/\alpha+1)}{\Gamma(\lambda-|m|/\alpha+1)}P^{-|m|/\alpha}_{\lambda}(\cos\theta)^{2}  \Big\{\frac{ \xi I_{\lambda+1/2}(n \kappa \xi)K_{\lambda+1/2}(n \kappa \xi)}{M^{2}(\eta+1)^{2}}-\gamma^{(0)}_{n\lambda}-\gamma^{(1)}_{n\lambda}-\gamma^{(2)}_{n\lambda}\Big\}  \nonumber\\
 \Sigma_{3}&=\sum_{l=0}^{\infty}
 \sum_{m=-l}^{l}(2 \lambda +1)\frac{\Gamma( \lambda+|m|/\alpha+1)}{\Gamma(\lambda-|m|/\alpha+1)}P^{-|m|/\alpha}_{\lambda}(\cos\theta)^{2} \Big\{ \frac{P_{\lambda}(\eta) Q_{\lambda}(\eta)}{M}-\beta^{(0)}_{0\lambda}-\beta^{(1)}_{0\lambda}-\beta^{(2)}_{0\lambda}\Big\} .
\end{align}
We have re-written the $l$, $m$ sum and swapped the order of summation with the $n$-sum which presents no problem for these sums since they are all rapidly and absolutely convergent. In fact, the summand of $\Sigma_{1}$ and $\Sigma_{2}$ are $O(l/(l^{2}+n^{2})^{7/2})$ for large $l$ and $n$. The summand of $\Sigma_{3}$ is $O(l^{-6})$ for large $l$. 

The slowly convergent term, $\Lambda$, is given by
\begin{align}
\label{eq:lambdasum}
\Lambda &= \sum_{l=0}^{\infty}
 \sum_{m=-l}^{l}(2 \lambda +1)\frac{\Gamma( \lambda+|m|/\alpha+1)}{\Gamma(\lambda-|m|/\alpha+1)}P^{-|m|/\alpha}_{\lambda}(\cos\theta)^{2} \Big\{ \frac{\beta^{(1)}_{0\lambda}}{2}+\frac{\beta^{(2)}_{0\lambda}}{2} +\sum_{n=1}^{\infty}\Big(\beta^{(1)}_{n\lambda}-\gamma^{(1)}_{n\lambda}+\beta^{(2)}_{n\lambda}-\gamma^{(2)}_{n\lambda} \Big) \Big\}. \nonumber\\
\end{align}
In order to speed the convergence of this sum, we can convert the $n$-sum here to an integral using a modified version of the Plana-Abel Sum Formula.

%-------------------PLANA-ABEL SUM FORMULA-------------------------------------------------------------

\section{Modified Plana-Abel Sum Formula}
 We begin by re-writing the difference of the WKB approximants of order $i$ ($i=1,2$) as~\cite{CDOW1}
 \begin{equation}
 \label{eq:wkbdiff}
 \beta_{n\lambda}^{(i)}-\gamma_{n\lambda}^{(i)}=\sum_{j=0}^{2 i}\frac{ 4^{2 i+1} C_{i j}(\eta)}{(\eta+1)^{4 i+2}} \frac{n^{2j}}{[\Omega^{2}+n^{2}]^{i+j+1/2}}
 \end{equation}
 where $\Omega$ is given by
 \begin{equation}
 \Omega=\frac{4(\lambda+1/2)(\eta^{2}-1)^{1/2}}{(\eta+1)^{2}}.
 \end{equation}
 and the coefficients $C_{ij}(\eta)$ are tabulated in Table 1.
 
 \begin{table}[htdp]
 \caption{The coefficients $C_{ij}(\eta)$ of the difference of the WKB approximants.}
\begin{center}
\begin{tabular}{| c || c | c | c | c | c |} \hline
$C_{ij}(\eta)$  &  0   &    1   &    2     &     3    &     4    \\
\hline\hline
1 & $\frac{1}{16 M}$ & $\frac{3}{8 M} (2\eta-3)$  &  $\frac{-5}{16 M} (4\eta-5)$ &\phantom{\Big|} & \\
%& & & & & \\
\hline
2 & $\frac{1}{256 M}(16\eta^{2}+11)$  &  $\frac{5}{16 M}(12\eta^{3}-24\eta^{2}$  &  $\frac{-35}{128 M}(64\eta^{3}-177\eta^{2}$  &   $\frac{231}{64 M}(14\eta^{3}-43\eta^{2}$  &   $\frac{-1155}{256 M}(8\eta^{3}-26\eta^{2}$ \\
    &    &$+14\eta-31)$     &  $+204\eta-122)$    & $+52\eta-26)$   &   $+32\eta-15)$\phantom{\big|}  \\
   
    \hline

\end{tabular}
\end{center}
\label{default}
\end{table}%

 We now convert the $n$-sum above to an integral using a modification of the Plana-Abel sum formula \cite{Fialkovsky, CDOW1}. In our particular case, we have
 \begin{equation}
 \label{eq:planasumformula}
 \sum_{n=1}^{\infty} \frac{n^{2j}}{[\Omega^{2}+n^{2}]^{i+j+1/2}}=-\frac{\delta_{j0}}{2 \Omega^{2 i+1}} +\int_{0}^{\infty}\frac{n^{2 j}}{[\Omega^{2}+n^{2}]^{i+j+1/2}} \mathrm{d}n+2(-1)^{i+j}\frac{\sqrt{\pi}}{\Gamma(i+j+1/2)}\int_{\Omega}^{\infty}\frac{h^{(i+j)}(s)}{(s-\Omega)^{1/2}} \mathrm{d}s 
 \end{equation}
where
\begin{equation}
h(s) = \frac{(s e^{i \pi/2})^{2 j}}{(s+\Omega)^{i+j+1/2}}\frac{1}{e^{2 \pi s}-1}.
\end{equation}
Furthermore, the $n$-integration of Eq.(\ref{eq:planasumformula}) can be done explicitly using
\begin{equation}
\label{eq:nintegration}
\int_{0}^{\infty}\frac{n^{2 j}}{[\Omega^{2}+n^{2}]^{i+j+1/2}} \mathrm{d}n=\frac{1}{\Omega^{2 i}}\frac{\Gamma(i)\Gamma(j+1/2)}{2\Gamma(i+j+1/2)}.
\end{equation}
Substituting Eq.(\ref{eq:nintegration}) into Eq.(\ref{eq:planasumformula}), we get
\begin{align}
\sum_{n=1}^{\infty} \big(\beta^{(i)}_{n\lambda}-\gamma^{(i)}_{n\lambda}\big)=-\frac{4^{2 i+1} C_{i0}(\eta)}{2 (\eta+1)^{4 i+2} \Omega^{2 i+1}} +\frac{4^{2 i+1}}{(\eta+1)^{4 i+2}}\sum_{j=0}^{\infty} \frac{C_{ij}(\eta)}{\Omega^{2 i}}\frac{\Gamma(i)\Gamma(j+1/2)}{2\Gamma(i+j+1/2)} \nonumber\\
+\frac{4^{2 i+1}}{(\eta+1)^{4 i+2}}\sum_{j=0}^{2 i} \frac{2 C_{ij}(\eta) (-1)^{i+j}\sqrt{\pi}}{\Gamma(i+j+1/2)}\int_{\Omega}^{\infty}\frac{h^{(i+j)}(s)}{(s-\Omega)^{1/2}} \mathrm{d}s 
\end{align}
Using the coefficients of Table 1, the first term here is conveniently just $- \beta_{0\lambda}/2$, while the second term here vanishes for $i=1,2$,  that is,
\begin{equation}
\sum_{j=0}^{2 i}\frac{C_{ij}(\eta) \Gamma(j+1/2)}{\Gamma(i+j+1/2)}=0\qquad\text{for} \quad i=1,2.
\end{equation}

Finally, we arrive at the following expression for the $n$-sum of the difference of the WKB approximants of $i^{th}$ order:
\begin{equation}
\label{eq:nsumtointegral}
\sum_{n=1}^{\infty} \big(\beta^{(i)}_{n\lambda}-\gamma^{(i)}_{n\lambda}\big)=-\frac{\beta^{(i)}_{0\lambda}}{2}+\frac{4^{2 i+1}}{(\eta+1)^{4 i+2}}\sum_{j=0}^{2 i} \frac{2 C_{ij}(\eta) (-1)^{i+j}\sqrt{\pi}}{\Gamma(i+j+1/2)}\int_{\Omega}^{\infty}\frac{h^{(i+j)}(s)}{(s-\Omega)^{1/2}} \mathrm{d}s 
\end{equation}

Returning now to our expression for $\Lambda$, we substitute (\ref{eq:nsumtointegral}) into (\ref{eq:lambdasum}). Remarkably, the $n=0$ terms cancel, leaving only the rapidly convergent integrals of Eq.(\ref{eq:nsumtointegral}), \textit{viz.},
\begin{equation}
\label{eq:lambdaintegral}
\Lambda = \sum_{m=-\infty}^{\infty}
 \sum_{l=|m|}^{\infty}(2 \lambda +1)\frac{\Gamma( \lambda+|m|/\alpha+1)}{\Gamma(\lambda-|m|/\alpha+1)}P^{-|m|/\alpha}_{\lambda}(\cos\theta)^{2} \Big\{ \sum_{i=1}^{2}\sum_{j=0}^{2 i}\frac{4^{2 i+1}}{(\eta+1)^{4 i+2}}\frac{2 C_{ij}(\eta) (-1)^{i+j}\sqrt{\pi}}{\Gamma(i+j+1/2)}\int_{\Omega}^{\infty}\frac{h^{(i+j)}(s)}{(s-\Omega)^{1/2}} \mathrm{d}s \Big\}.
\end{equation}
\end{widetext}

%------------------------NUMERICAL EVALUATION--------------------------------------------------------------------

\section{Numerical Evaluation of $\langle \hat{\varphi}^{2} \rangle_{ren}$}
We have shown that we can write the renormalized vacuum polarization outside the horizon as an analytic part Eq.~(\ref{eq:phi2analytic}), a regular integral part Eq.~(\ref{eq:phi2int}) and a contribution coming from the mode-sum 
(Eqs.~(\ref{eq:phi2sumwkb})-(\ref{eq:sigmaterms}) and Eq.~(\ref{eq:lambdaintegral})). In addition, we have calculated elsewhere~\cite{CSHorizon} the vacuum polarization as an analytic expression on the horizon,
\begin{equation}
\label{eq:phi2horizon}
 \langle \hat{\varphi}^{2} \rangle_{ren}^{horizon} = \frac{1}{192\pi^{2} M^{2}}\Big(1+\frac{(1-\alpha^{2})}{\alpha^{2}\sin^{2}\theta}\Big).
  \end{equation}
Combining these results gives the renormalized vacuum polarization on the entire spacetime region of interest. 

We now turn to a specific example. Thus far, we have described the calculation of $\langle \hat{\varphi}^{2} \rangle_{ren}$ on the exterior of the Schwarzschild spacetime threaded by a cosmic string from the horizon out to infinity. The results we shall show, however, are for a Schwarzschild blackhole threaded by an infinite cosmic string inside a spherical box. The reasons for this are twofold: Firstly, it is numerically easier. Rather than evaluating numerical radial modes over the entire radial range, we are now limited to a finite range between the horizon and the boundary. Secondly, this calculation is motivated by the analagous Kerr renormalization. In this case, one cannot define a Hartle-Hawking vacuum everywhere on the exterior of the black hole. However, in order to determine the vacuum polarization on a state that possesses the defining features of a Hartle-Hawking vacuum, one must put in a spherical mirror inside the speed of light surface \cite{DuffyOttewill}. In any case, as a check of our method, we have also run the calculation where we take the boundary out to a very large radius, which yields the Candelas-Howard result \cite{Candelas:1984pg} in the $\alpha \rightarrow 1$ limit.

For the results shown, we have taken units where the mass $M=1$, the azimuthal deficit $\alpha=0.95$ and the boundary imposes Dirichlet boundary conditions at $\eta_{b}=3$ (or $r_{b}= 4 M$). We take an $(r,\theta)$ grid that consists of 50 radial points and 70 angular points.This boundary condition requires that the Green's function vanish on the mirror, implying that the outer radial function becomes
\begin{align}
 q_{n\lambda}^{(b)}(\eta)= q_{n\lambda}(\eta)
 -\frac{q_{n\lambda}(\eta_{b})}{p_{n\lambda}(\eta_{b})}{p_{n\lambda}(\eta)}.
\end{align}
On the horizon, we now have
\begin{widetext}
\begin{equation}
 \langle \hat{\varphi}^{2} \rangle_{ren}^{horizon,box} = \frac{1}{192\pi^{2} M^{2}}\Big(1+\frac{(1-\alpha^{2})}{\alpha^{2}\sin^{2}\theta}\Big)  -\frac{1}{32\pi^{2} M^{2}\alpha}\sum_{l=0}^{\infty}\sum_{m=-l}^{l}(2 \lambda +1)\frac{\Gamma( \lambda+|m|/\alpha+1)}{\Gamma(\lambda-|m|/\alpha+1)}P^{-|m|/\alpha}_{\lambda}(\cos\theta)^{2}\frac{Q_{\lambda}(\eta_{b})}{P_{\lambda}(\eta_{b})}.
 \end{equation}
\end{widetext}
The sum here converges rapidly (about 12 decimal places after only 20 $l$-modes). 

With the exception of the boundary terms above, the calculation inside a spherical box remains unchanged. The calculation of $\langle \hat{\varphi}^{2} \rangle_{analytic}$ is trivial and requires no discussion. Similarly, the calculation of $\langle \hat{\varphi}^{2} \rangle_{int}$ is numerically evaluated over the grid without difficulty. As we have mentioned the sums of $\Sigma$ (Eq.(\ref{eq:sigmaterms})) are $O(l/(n^{2}+l^{2})^{7/2})$ and are therefore rapidly convergent. Close to the horizon and the boundary, however, this behaviour doesn't become apparent in $\Sigma_{1}$ and $\Sigma_{3}$ until the higher $l$ and $n$-modes, compared with the interior region of interest. For this reason, after 100 $l$-modes and 20 $n$-modes, we see convergence to about 6 decimal places near the horizon and the boundary and about 10 decimal places in the interior for $\Sigma_{1}$ and $\Sigma_{2}$. We see convergence to at least 10 decimal places everywhere for $\Sigma_{2}$ for 100 $l$-modes and 20 $n$-modes. The expression $\Lambda$ (Eq.(\ref{eq:lambdaintegral})) is exponentially convergent. Again, this convergence does not become apparent until the higher $l$-modes near the horizon. We observe convergence to at least 10 decimal places everywhere in the region of interest by taking 30 $l$-modes near the horizon and 10 $l$-modes away from the horizon. 

\begin{figure}
\centering
\includegraphics[width=8cm]{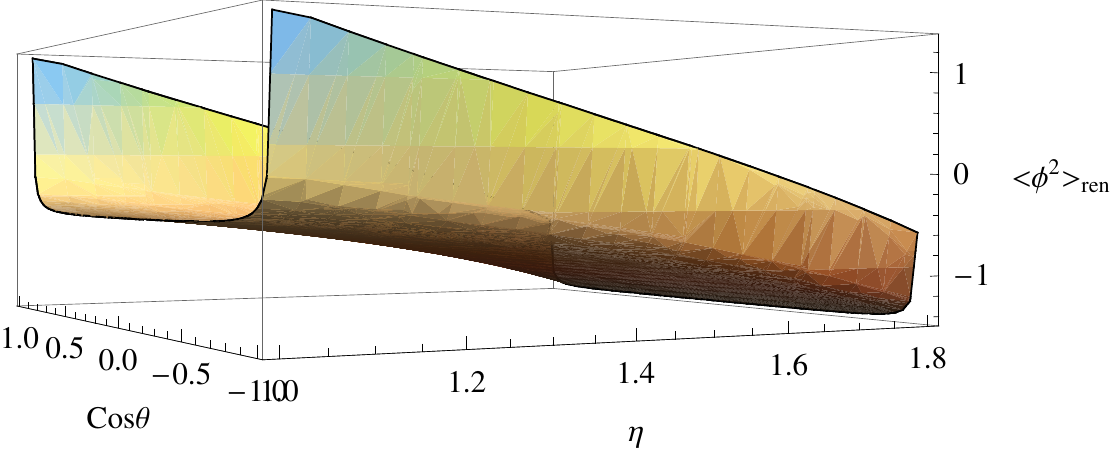}
\caption{{Plot of $(8 \pi M)^{2}\langle \hat{\varphi}^{2} \rangle_{ren}$}. It is clear from the figure that the vacuum polarization diverges at the poles, as we would expect since there is a curvature singularity there. }
\label{fig:plotphi2ren3D}
\end{figure} 

%\begin{figure}
%\centering
%\includegraphics[width=8cm]{plot2Dphi2ren1.pdf}
%\caption{\emph{Plot of $(8 \pi M)^{2}\langle \hat{\varphi}^{2} \rangle_{ren}$ for particular values of $\cos\theta$.}}
%\label{fig:plot2Dphi2ren1}
%\end{figure} 

\begin{figure}
\centering
\includegraphics[width=8cm]{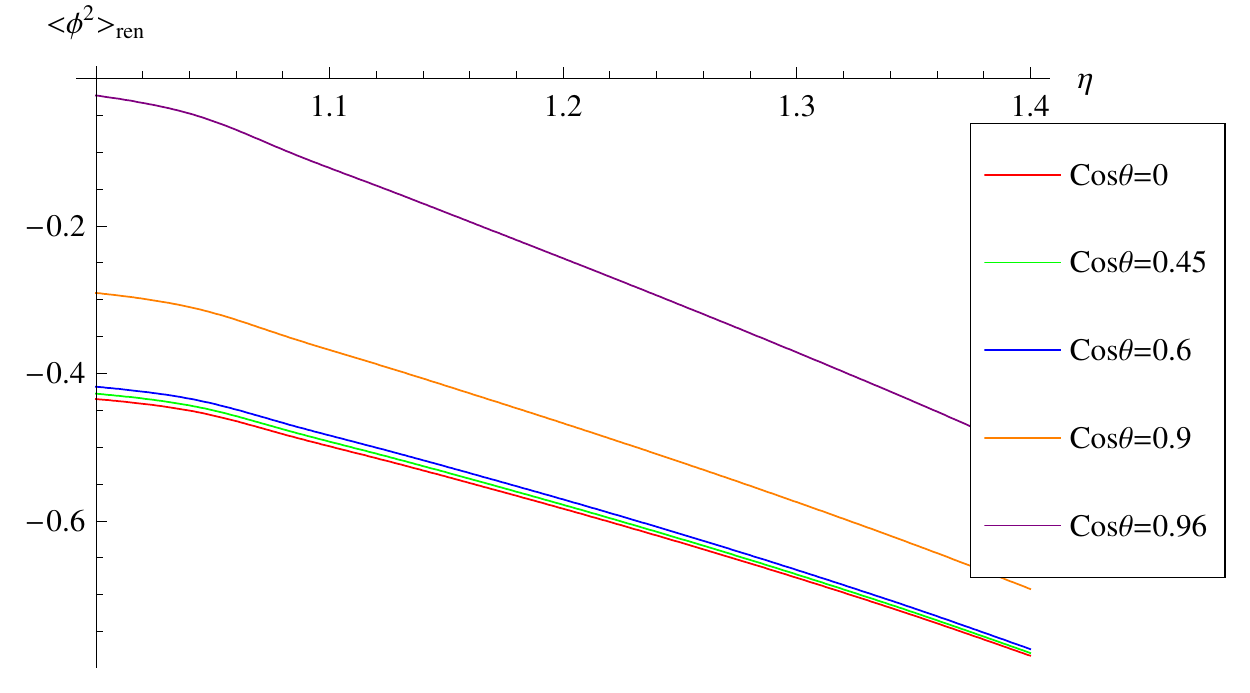}
\caption{{Plot of $(8 \pi M)^{2}\langle \hat{\varphi}^{2} \rangle_{ren}$ for particular values of $\cos\theta$.} We observe that the effect of the string only becomes important very close to the string. This is due to $\alpha$ being close to~1.}
\label{fig:plot2Dphi2ren2}
\end{figure} 

\begin{figure}
\centering
\includegraphics[width=8cm]{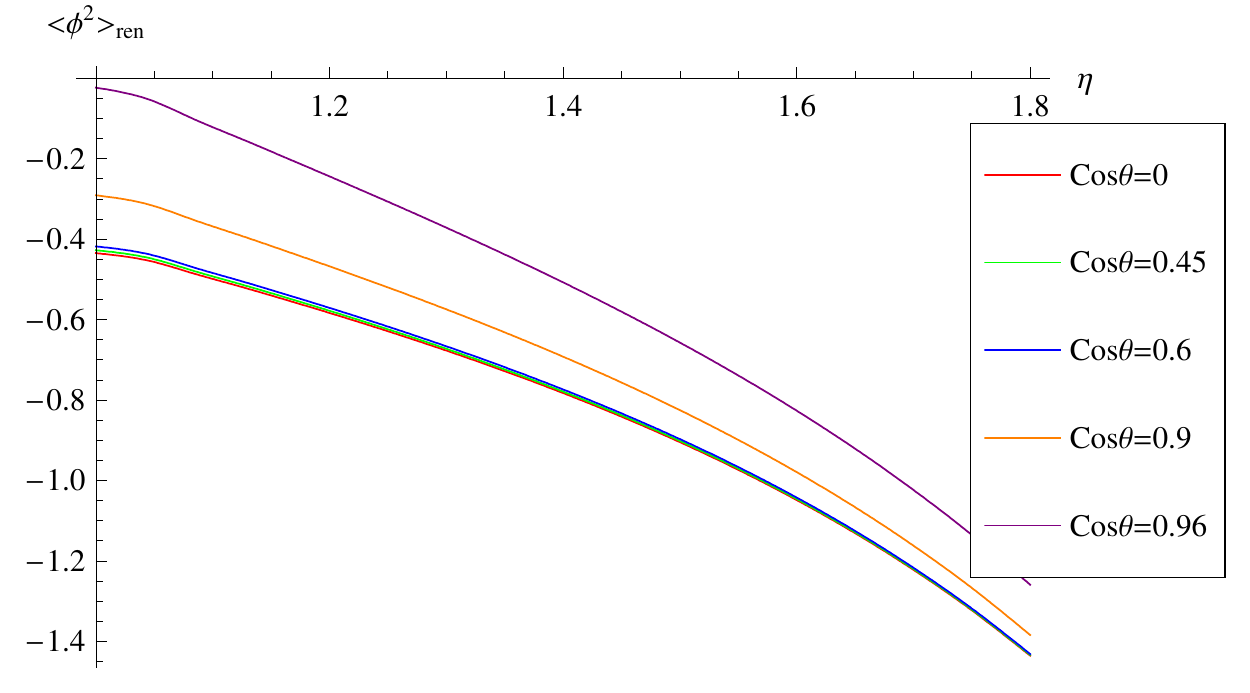}
\caption{{Plot of $(8 \pi M)^{2}\langle \hat{\varphi}^{2} \rangle_{ren}$ for particular values of $\cos\theta$.}}
\label{fig:plot2Dphi2ren3}
\end{figure} 

\begin{figure}
\centering
\includegraphics[width=8cm]{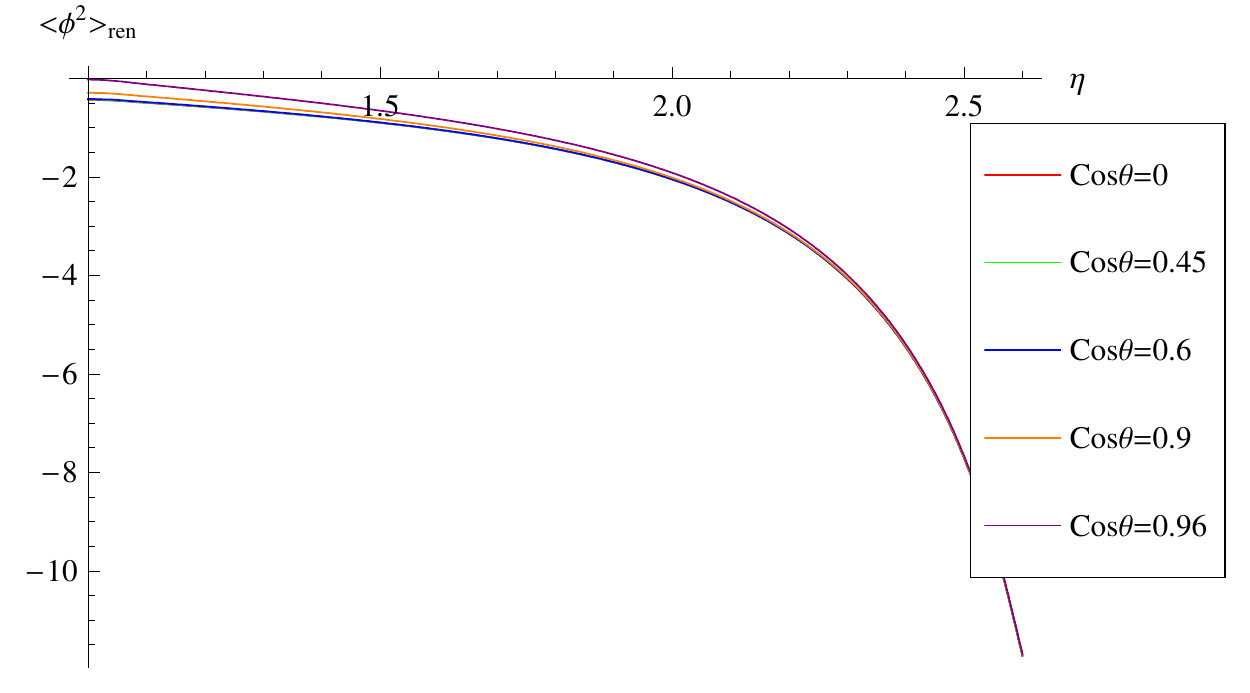}
\caption{{Plot of $(8 \pi M)^{2}\langle \hat{\varphi}^{2} \rangle_{ren}$ for particular values of $\cos\theta$.}}
\label{fig:plot2Dphi2ren4}
\end{figure} 

We see from the graphs that the Casimir effect, arising from the presence of the boundary at $\eta_{b}=3$, actually dominates when the boundary is this close to the horizon and becomes divergent as we approach the boundary. We also note that the vacuum polarization increases (becomes more positive) as we get closer to the poles. Again, we would expect this since there is a curvature singularity on the poles due to the cosmic string. In this particular case, since the azimuthal deficit is so small, the effect of the cosmic string is almost negligible until we get very close to the poles, responsible for the sharp divergence in Fig.~\ref{fig:plotphi2ren3D}. This graph would be more rounded and smooth for smaller values of $\alpha$ (corresponding to larger azimuthal deficits). We see from Fig.~\ref{fig:plot2Dphi2ren2} that the cosmic string has little effect on the vacuum polarization when we are not close to the poles. The $\theta$ dependence diminishes the further we are from the horizon, i.e., when we are not close to the pole of the black hole, the effect of the cosmic string diminishes as we move further from the horizon. This is evident by comparing Fig.~\ref{fig:plot2Dphi2ren3} and Fig.~\ref{fig:plot2Dphi2ren4}, since we can see the merging of the graphs as we move away from the horizon. The Casimir divergence due to the boundary is also clear from  Fig.~\ref{fig:plot2Dphi2ren4}.

\begin{figure}
\centering
\includegraphics[width=8cm]{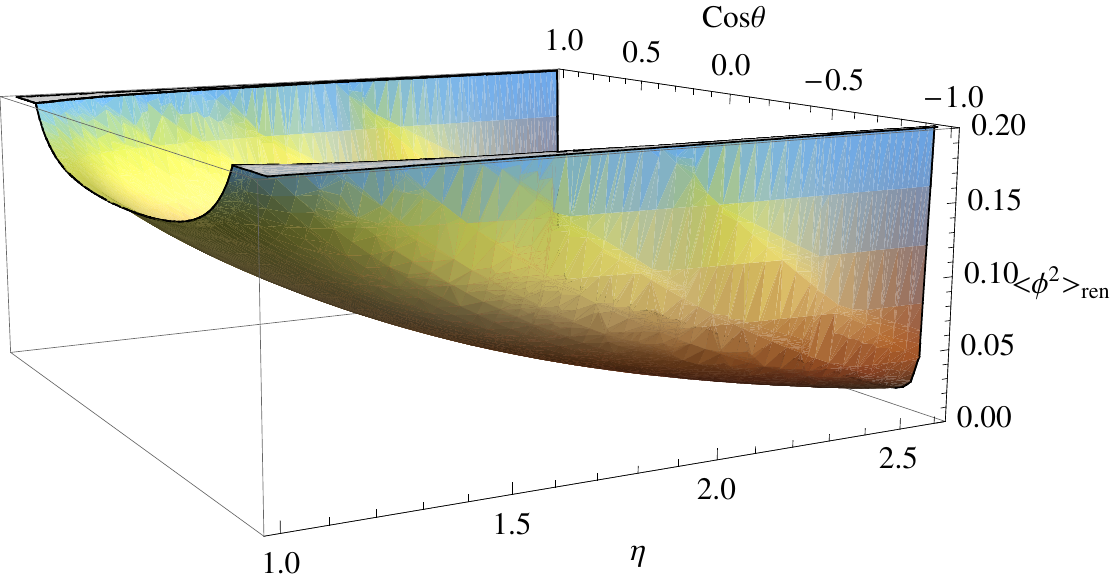}
\caption{{Plot of $(8 \pi M)^{2}(\langle \hat{\varphi}^{2} \rangle_{ren}-\langle \hat{\varphi}^{2} \rangle_{div}^{Casimir})$.}}
\label{fig:phi2renminuscasimir}
\end{figure}

\begin{figure}
\centering
\includegraphics[width=8cm]{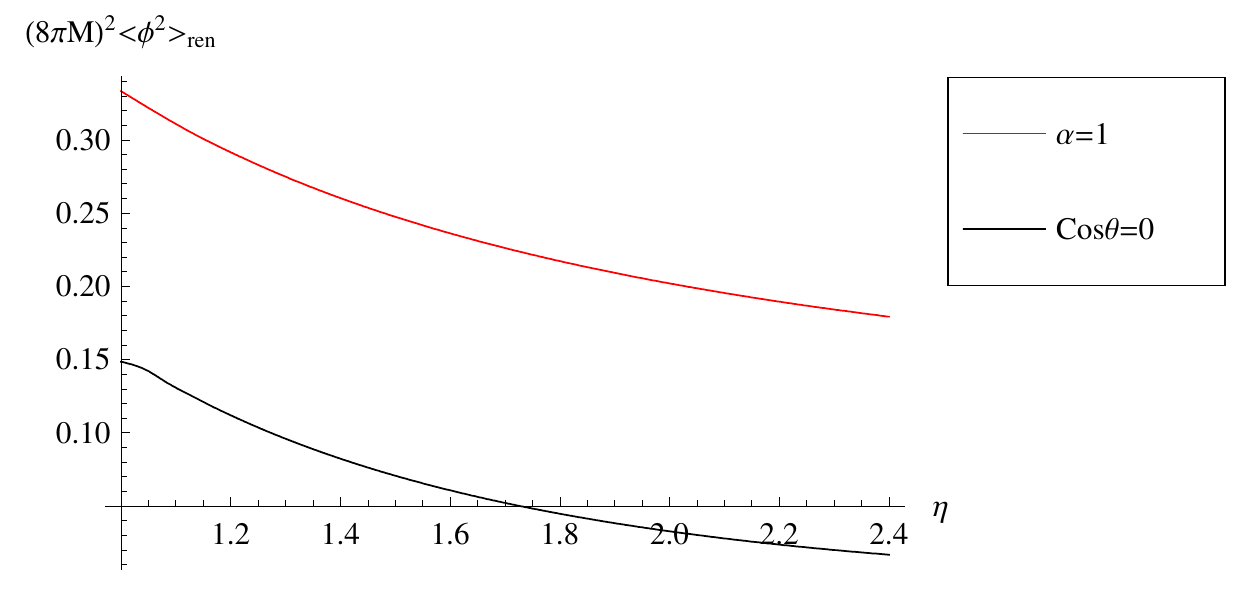}
\caption{{Plot of $(8 \pi M)^{2}(\langle \hat{\varphi}^{2} \rangle_{ren}-\langle \hat{\varphi}^{2} \rangle_{div}^{Casimir})$ and $(8 \pi M)^{2}\langle \hat{\varphi}^{2} \rangle_{ren}^{\alpha\rightarrow 1}$. }}
\label{fig:phi2renschwarzschild}
\end{figure}

The dominance of the Casimir effect serves to obscure the behaviour one would expect to see in the absence of a boundary. However, we may subtract the Casimir divergence by doing a DeWitt expansion of the heat kernel about the boundary~\cite{McAvityOsborn}. For our particular geometry of spacetime and surafce, we find that the expected 
divergence from the curved boundary is given by
\begin{equation}
\langle \hat{\varphi}^{2} \rangle_{div}^{Casimir} = -\frac{1}{32 \pi^{2} M^{2}} \Big(\frac{1}{(\eta_{b}-\eta)^{2}}+\frac{1}{12(\eta_{b}-\eta)}\Big).
\end{equation}
Subtracting these divergent terms provides a clear picture (modulo finite boundary effects) of $\langle \hat{\varphi}^{2} \rangle_{ren}$ in the absence of a boundary without evaluating the mode-sum over the entire exterior region of the black hole. In particular, Fig.~\ref{fig:phi2renminuscasimir} possesses the general features of the renormalized vacuum polarization without a boundary. This is very evident from Fig.~\ref{fig:phi2renschwarzschild} where we have plotted $\langle \hat{\varphi}^{2} \rangle_{ren}$ for the Schwarzschild case ($\alpha \rightarrow 1$, i.e. no cosmic string) using the data tabulated by Candelas and Howard~ \cite{Candelas:1984pg} with the plot of $\langle \hat{\varphi}^{2} \rangle_{ren}-\langle \hat{\varphi}^{2} \rangle_{div}^{Casimir}$ on the equatorial plane. The profiles are clearly identical, differing only by a constant finite amount, though this finite difference is constant only in the equatorial case. Nevertheless, it is clear that the essential features have been captured without the computationally intensive evaluation of the mode-sum out to `infinity', or in reality,  some radius very far from the horizon.
%\begin{widetext}

%\end{widetext}

%--------------------------------CONCLUSIONS------------------------------------------------------------

\section{Conclusions}

With the exception of Minkowski spacetime threaded by a cosmic string, explicit 
calculation of  physically important renormalized expectation values in the axially symmetric geometry has proved elusive, though it is worth mentioning that Davies and Sahni \cite{DaviesSahni} have calculated the vacuum polarization on the horizon of a Schwarzschild black hole threaded by a cosmic string for the restricted range $(1/\alpha)$ an integer, a result which we generalized to all $\alpha$ in \cite{CSHorizon}. Our calculation in this paper represents the first renormalization of the vacuum polarization on the exterior region of an axially symmetric black hole spacetime. Most importantly, we have renormalized without using properties of the special functions, (such as Addition Theorems for the Legendre functions, for example), but rather we have obtained useful summation formulae over the required mode functions that express geometrical singularities as non-convergent mode-sums. Moreover, we have appealed to the Hadamard singularity structure on a dimensionally reduced spacetime in order to guarantee the convergence of our renormalized expectation value. Our resultant expressions were rapidly convergent and relatively easy to compute numerically. This approach makes this method attractive in the renormalization process on the Kerr black hole, where we do not have the luxury of Addition Theorems. 

In addition, we have elucidated some points on how the direction in which we choose to point-split affects the order of our summation. We showed that a temporal splitting enforces an inner $l$, $m$-sum and while for azimuthal splitting an inner $n$-sum is required. To the best of our knowledge, this subtlety has not previously been noted, perhaps because it has most often been convenient to split in the temporal direction in the spherically symmetric case; we became aware of it having, without due care in this direction,  obtained different results for the same expression using azimuthal and temporal separation. Understanding of such issues is of key importance to extend our results to the Kerr space-time, since provisional calculations suggest that it is more convenient in that case to split in the azimuthal direction, as we have done in this paper. 

%------------------------------ACKNOWLEDGEMENTS---------------------------------------------------

\section{Acknowledgments}
We would like to thank Elizabeth Winstanley for her careful reading of the manuscript and for her helpful suggestions.

PT is supported by the Irish Research Council for Science, Engineering and Technology, funded by the National
Development Plan.

%----------------------------APPENDIX--------------------------------------------------------------

\section*{Appendix}

In this appendix, we illustrate how to find the Hadamard divergence of the $l$, $m$ mode-sum of the Green's function. We first note that since we are only interested in the divergent part, the order of summation is irrelevant since different orders result in a finite discrepancy. Separating out the temporal dependence by writing
\begin{equation}
\label{eq:gn}
G(x,x') = \frac{T}{4\pi}\sum_{n=-\infty}^{\infty}e^{in\kappa(\tau-\tau')} G_{n}(\mathbf{x},\mathbf{x'}),
\end{equation}
and substituting into the wave equation, we find
\begin{align}
&\Big(\frac{\partial}{\partial r}(r^{2}-2 M r)\frac{\partial}{\partial r} +\frac{1}{\sin\theta}\frac{\partial}{\partial\theta}\sin\theta \frac{\partial}{\partial\theta}
 +\frac{1}{\alpha^{2} \sin^{2}\theta}\frac{\partial^{2}}{\partial\phi^{2}}
 \nonumber\\
&\qquad\qquad-\frac{n^{2}\kappa^{2}r^{4}}{r^{2}-2 M r} \Big)G_{n} = -\frac{\delta(\mathbf{x}-\mathbf{x'})}{\alpha \sin\theta}. 
\end{align}
Multiplying across by $(1-2M/r)^{-1}$, this equation is a 3-dimensional Laplace equation with potential 
\begin{equation}
(\nabla^{2}-V)G_{n}(\mathbf{x},\mathbf{x'})=-g_{3}^{-1/2} \delta(\mathbf{x}-\mathbf{x'})
\end{equation}
where $\nabla^{2}$ is the invariant Laplacian on the 3-metric
\begin{equation}
\label{eq:3metric}
\mathrm{d}s_{3}^{2}=\mathrm{d}r^{2} +(r^{2}-2 M r) \mathrm{d}\theta^{2} +(r^{2}-2 M r)\alpha^{2}\sin^{2}\theta \mathrm{d}\phi^{2}
\end{equation}
where the potential is given by
\begin{equation}
V=\frac{n^{2}\kappa^{2} r^{6}}{(r^{2}-2 M r)^{2}}.
\end{equation}

An important result from Hadamard \cite{Hadamard} is that the singularity structure of the Green's function is independent of the potential and depends only on the principal part of the wave operator. In 3 dimensions, the Hadamard form is 
\begin{equation}
G_{n}(\mathbf{x},\mathbf{x'})\sim\frac{U(\mathbf{x},\mathbf{x'})}{(2 \sigma(\mathbf{x},\mathbf{x'}))^{1/2}} +W(\mathbf{x},\mathbf{x'})
\end{equation}
where $2\sigma$ is the square of the geodesic distance on the 3-metric (\ref{eq:3metric}) and $U(\mathbf{x},\mathbf{x'})$, $W(\mathbf{x},\mathbf{x'})$ are regular biscalars. One can expand $U$ and $W$ in terms of $\sigma$ \cite{Decanini:2008}, where to the required order we have
\begin{align}
U(\mathbf{x},\mathbf{x'}) &= 1+O(\sigma) \nonumber\\
W(\mathbf{x},\mathbf{x'}) &=O(1).
\end{align}
For azimuthal separation, we have
\begin{equation}
2 \sigma = (r^{2}-2 M r)\alpha^{2}\sin^{2}\theta \Delta\phi^{2} +O(\Delta\phi)^{4}
\end{equation}
so that the Green's function at point separated in this way is
\begin{align}
\label{eq:gndiv}
&G_{n}(r,\theta,\Delta\phi)= \frac{1}{ (r^{2}-2 M r)^{1/2}\alpha \sin\theta \Delta\phi} + O(1) \nonumber\\
&\qquad=\frac{1}{ M(\eta^{2}-1)^{1/2}\alpha \sin\theta \Delta\phi} + O(1) . 
\end{align}
%We have used the radial transformation (\ref{eq:eta}) to obtain the the final form above. 
We also have the equivalent mode-sum expression for $G_{n}$, obtained by comparing Eq.~(\ref{eq:gn}) with 
Eq.~(\ref{eq:greensfn}). Equating this mode-sum with expression (\ref{eq:gndiv}) gives
\begin{widetext}
\begin{equation}
\label{eq:modesumdiv}
\sum_{l=0}^{\infty}\sum_{m=-l}^{l} e^{im\Delta\phi} (2\lambda+1)\frac{\Gamma(\lambda+|m|/\alpha+1)}{\Gamma(\lambda-|m|/\alpha+1)}P_{\lambda}^{-|m|/\alpha}(\cos\theta)^{2}\chi_{n\lambda}(\eta,\eta) =\frac{1}{ M(\eta^{2}-1)^{1/2}\alpha \sin\theta \Delta\phi} + O(1)
\end{equation}
In addition, we have derived elsewhere~\cite{CSHorizon} the following result for these Legendre Functions
\begin{align}
\frac{1}{[2(1-\cos\alpha\gamma)]^{1/2}}= \frac{1}{\alpha}\sum_{l=0}^{\infty}\sum_{m=-l}^{l} e^{im\Delta\phi}\frac{\Gamma(\lambda+|m|/\alpha+1)}{\Gamma(\lambda-|m|/\alpha+1)}P_{\lambda}^{-|m|/\alpha}(\cos\theta)P_{\lambda}^{-|m|/\alpha}(\cos\theta') \nonumber\\
-\frac{1}{2\pi\alpha} \int_{0}^{\infty}\frac{F_{\alpha}(u,\phi-\phi')}{[2(1-\cos\theta\cos\theta'+\sin\theta\sin\theta'\cosh u)]^{1/2}} \mathrm{d}u.
\end{align}
where
\begin{equation}
\cos\alpha\gamma = \cos\theta\cos\theta'+\sin\theta\sin\theta'\cos\alpha\Delta\phi
\end{equation}
and
\begin{equation}
F_{\alpha}(u,\Psi) = \frac{sin(\Psi-\pi/\alpha)}{\cosh(u/\alpha)-\cos(\Psi-\pi/\alpha)} - \frac{sin(\Psi+\pi/\alpha)}{\cosh(u/\alpha)-\cos(\Psi+\pi/\alpha)}.
\end{equation}
This formula relates a geometrical Hadamard singularity to a non-convergent mode-sum over the Legendre functions, valid for $1/2<\alpha\leq1$. The integral term here is regular everywhere (apart from the poles where there is a curvature singularity) and vanishes for $\alpha = 1$.  For small $\phi$ separations, this becomes
\begin{equation}
\frac{1}{\alpha\Delta\phi \sin\theta } =\frac{1}{\alpha}\sum_{l=0}^{\infty}\sum_{m=-l}^{l} e^{im\Delta\phi}\frac{\Gamma(\lambda+|m|/\alpha+1)}{\Gamma(\lambda-|m|/\alpha+1)}P_{\lambda}^{-|m|/\alpha}(\cos\theta)^{2}+O(1).
\end{equation}
Dividing across by $M(\eta^{2}-1)^{1/2}$ and substituting into Eq.(\ref{eq:modesumdiv}) yields
\begin{equation}
\label{eq:hadamardsing}
\sum_{l=0}^{\infty}\sum_{m=-l}^{l} e^{im\Delta\phi} (2\lambda+1)\frac{\Gamma(\lambda+|m|/\alpha+1)}{\Gamma(\lambda-|m|/\alpha+1)}P_{\lambda}^{-|m|/\alpha}(\cos\theta)^{2}\Big(\chi_{n\lambda}(\eta,\eta)-\frac{1}{(2\lambda+1)M(\eta^{2}-1)^{1/2}}\Big)=O(1),
\end{equation}
Therefore, we have shown that the subtraction term here captures the divergence of the 3D Green's function $G_{n}$ as a mode-sum. 

In a completely analagous way, one can associate the divergence of the $l$, $m$ mode-sum in the subtraction terms of
 Eq.~(\ref{eq:phi2sum}) with the divergence of 3D Green's function with a particular potential. The key point here is that this 3D divergence structure is captured entirely by the $n=0$ term since all $n\ne0$ terms differ only by a  potential and so do not affect the singularity~\cite{Hadamard}. Thus, for a given $n$, the singular part of the $l$, $m$ sum of 
Eq.~(\ref{eq:phi2sum}) is captured by the $n=0$ term
\begin{equation}
\sum_{l=0}^{\infty}\sum_{m=-l}^{l} e^{im\Delta\phi} (2\lambda+1)\frac{\Gamma(\lambda+|m|/\alpha+1)}{\Gamma(\lambda-|m|/\alpha+1)}P_{\lambda}^{-|m|/\alpha}(\cos\theta)^{2}\frac{\xi}{M^{2}(\eta+1)^{2}(2\lambda+1)}
\end{equation}
Comparing this to the subtraction term of Eq.~(\ref{eq:hadamardsing}) we see that we can guarantee the regularity of the $l$, $m$ sum for a given $n$ by choosing $\xi$ according to:
\begin{equation}
\xi=M\frac{(\eta+1)^{2}}{(\eta^{2}-1)^{1/2}}.
\end{equation}

\end{widetext}
%---------------------------------------------------------------------------------------------------------

%%%%%%%%%%%%%%%%%%%%%%%%%%%%%%%%%%%%%%%%%%%%%%%%%%%%%%%%%%%%%%%%%%%%%%%%%%%%%%%%%%%%%%%

\bibliographystyle{apsrev}

\end{document}